\input{aipcheck.tex}
\documentclass{aipproc}
\layoutstyle{6x9}

\begin{document}
\def\dsl{\,\raise.15ex\hbox{/}\mkern-13.5mu D} %this one can be subscripted
\def\xslash#1{{\rlap{$#1$}/}}
\def\openone{\leavevmode\hbox{\small1\kern-4.2pt\normalsize1}}
\def\ssqr#1#2{{\vbox{\hrule height #2pt
      \hbox{\vrule width #2pt height#1pt \kern#1pt\vrule width #2pt}
      \hrule height #2pt}\kern- #2pt}}
\def\sqr{\mathchoice\ssqr8{.4}\ssqr8{.4}\ssqr{5}{.3}\ssqr{4}{.3}}
\def\onedot{\makebox(0,0){$\scriptstyle 1$}}
\def\twodot{\makebox(0,0){$\scriptstyle 2$}}
\def\threedot{\makebox(0,0){$\scriptstyle 3$}}
\def\fourdot{\makebox(0,0){$\scriptstyle 4$}}
\def\Tr{{\rm Tr}}
\def\onebox{{\vbox{\hbox{$\sqr\thinspace$}}}}
\def\twobox{{\vbox{\hbox{$\sqr\sqr\thinspace$}}}}
\def\threebox{{\vbox{\hbox{$\sqr\sqr\sqr\thinspace$}}}}
\def\nbox{\hbox{$\sqr\sqr\sqr\sqr\raise2.7pt\hbox{$\,\cdot\cdot\cdot
\cdot\cdot\,$}\sqr\sqr\sqr\thinspace$}}
\def\nboxE{\vbox{\hbox{$\sqr\sqr\sqr\raise2.7pt\hbox{$\,\cdot\cdot\cdot
\cdot\cdot\,$}\sqr\sqr\sqr\sqr$}\nointerlineskip 
\kern-.2pt\hbox{$\sqr\sqr\sqr\raise2.7pt\hbox{$\,\cdot\cdot\cdot
\cdot\cdot\,$}\sqr$}}}
\def\nboxF{\vbox{\hbox{$\sqr\sqr\sqr\sqr\raise2.7pt\hbox{$\,\cdot\cdot\cdot
\cdot\cdot\,$}\sqr\sqr$}\nointerlineskip 
\kern-.2pt\hbox{$\sqr\sqr\sqr\sqr\raise2.7pt\hbox{$\,\cdot\cdot\cdot
\cdot\cdot\,$}\sqr$}}}

\title
[Topics on heavy baryon chiral perturbation theory in the large $N_c$
limit]{Topics on heavy baryon chiral perturbation theory in the large $N_c$
limit}

\author{Rub\'en Flores-Mendieta}{
  address={Instituto de F{\'\i}sica, Universidad Aut\'onoma de San Luis
  Potos{\'\i}, Alvaro Obreg\'on 64, San Luis Potos{\'\i}, S.L.P.\ 78000,
  Mexico},
}

\copyrightyear  {2001}

\begin{abstract}
We compute nonanalytical pion-loop corrections to baryon masses in a
combined expansion in chiral symmetry breaking and $1/N_c$, where $N_c$ is the
number of colors. Specifically, we compute flavor-$\bf 27$ baryon mass
splittings at leading order in chiral perturbation theory. Our results, at the
physical value $N_c=3$, are compared with the expressions obtained in
heavy baryon chiral perturbation theory with no $1/N_c$ expansion.
\end{abstract}

\date{\today}

\maketitle

\section{Introduction}

     Chiral perturbation theory has been a useful tool in the understanding
of low-energy QCD hadron dynamics. Its application to baryons through a new
formulation of the low-energy chiral effective Lagrangian in which the baryons
appear as heavy static fields, was first introduced in
Refs.~\cite{jm255,jm259}. The chiral Lagrangian thus obtained was used to
compute the leading nonanalytic in $m_s$ corrections to baryon axial
currents~\cite{jm255,jm259,bora}, masses, and non-leptonic
decays~\cite{jm255,jm259}, to name but a few. 

     Similarly, the $1/N_c$ expansion has proved to be useful in the analysis
of the spin-flavor structure of baryons in
QCD~\cite{carone,lmr,dm315,djm94,djm95}. Evidence for the predictions of
the $1/N_c$ expansion for baryons has been found in the analysis of
masses~\cite{djm94,djm95,jl}, magnetic moments~\cite{djm94,djm95,lmrw}, and
axial and vector currents~\cite{djm95,dai,fmjm}.

     The next natural step has been to combine both approaches so that
baryon matrix elements are obtained in a simultaneous chiral and $1/N_c$
expansion~\cite{djm94,lmr,luty,jen96,bed}. The resulting approach,
referred to as large-$N_c$ chiral perturbation theory, has proven to have a
significant predictive power in the analysis of the baryon mass
spectra~\cite{luty,bed,weise}, magnetic moments~\cite{lmrw} and axial
current~\cite{weise}.

     Here we present an explicit calculation of flavor-{\bf 27}
mass splittings of the octet and decuplet, which are calculable and
nonanalytic in the quark masses and baryon hyperfine mass splittings at
leading order in chiral perturbation theory. This analysis illustrates how to
implement the procedure.

\section{Baryons in chiral perturbation theory}

     The heavy baryon chiral Lagrangian can be constructed in terms of the
pion field $\Pi$, the baryon octet field $B_v$, and the baryon decuplet field
$T^\mu_{abc}$, where the $\Pi$ and $B_v$ fields are represented by $3 \times
3$ matrices and $T_{abc}^\mu$ is a Rarita-Schwinger field which contains both
spin-1/2 and spin-3/2 pieces and obeys the constraint $\gamma^\mu T_\mu = 0$
\cite{jm255,jm259}.

     The coupling of the pseudoscalar pion field with the baryon matter
fields occurs through the vector and axial vector combinations
$V^\mu = (1/2)(\xi \partial^\mu \xi^\dagger + \xi^\dagger \partial^\mu
\xi)$ and $A^\mu = (i/2)(\xi \partial^\mu \xi^\dagger -
\xi^\dagger \partial^\mu \xi)$, where $\xi = e^{i \Pi/f}$, $\Sigma = \xi^2 =
e^{2i \Pi/f}$, and $f \approx 93$ MeV is the pion decay constant. Further
considerations can be found in Refs.~\cite{jm255,jm259}.

     The most general Lagrangian at lowest order is
\begin{eqnarray}
{\cal L}_{\rm baryon} & = & i \, {\rm Tr} \, \bar B_v (v \cdot {\cal D})
B_v  - i \, \bar T_v^\mu (v \cdot {\cal D}) T_{v\mu} + \Delta \, \bar
T_v^\mu T_{v\mu} + 2 \, D \, {\rm Tr} \, \bar B_v S_v^\mu \{ A_\mu, B_v \}
\nonumber \\
&   & \mbox{} + 2 \, F \, {\rm Tr} \, \bar B_v S_v^\mu [ A_\mu, B_v ] + {\cal
C} \, ( \bar T_v^\mu A_\mu B_v + \bar B_v A_\mu T_v^\mu) + 2 \, {\cal H} \,
\bar T_v^\mu S_v^\nu A_\nu T_{v\mu} \, , 
\label{eq:efflag}
\end{eqnarray}
where $D$, $F$, $\cal C$, and $\cal H$ are the baryon-pion couplings and
$\Delta = m_T - m_B$ is the decuplet-octet mass difference. ${\cal L}_{\rm
baryon}$ describes massless pion fields interacting with degenerate SU(3)
multiplets of baryons.

     Three nonanalytic terms for the baryon masses are calculable, namely,
the ones which vary as $m_s^{3/2}$, $m_s^2 \ln{m_s}$, and $(\Delta) m_s
\ln{m_s}$. These contributions result from one-loop diagrams and have been
computed in Ref.~\cite{jen92}. Here we analyze the leading nonanalytic
contributions arising from the Feynman diagram displayed in Fig.~1. This
diagram involves $\pi$, $K$, and $\eta$ emission and reabsorption. The most
general form of this contribution can be written as
\begin{figure}\label{wavefunc} \resizebox{.4\textwidth}{!}
{\includegraphics{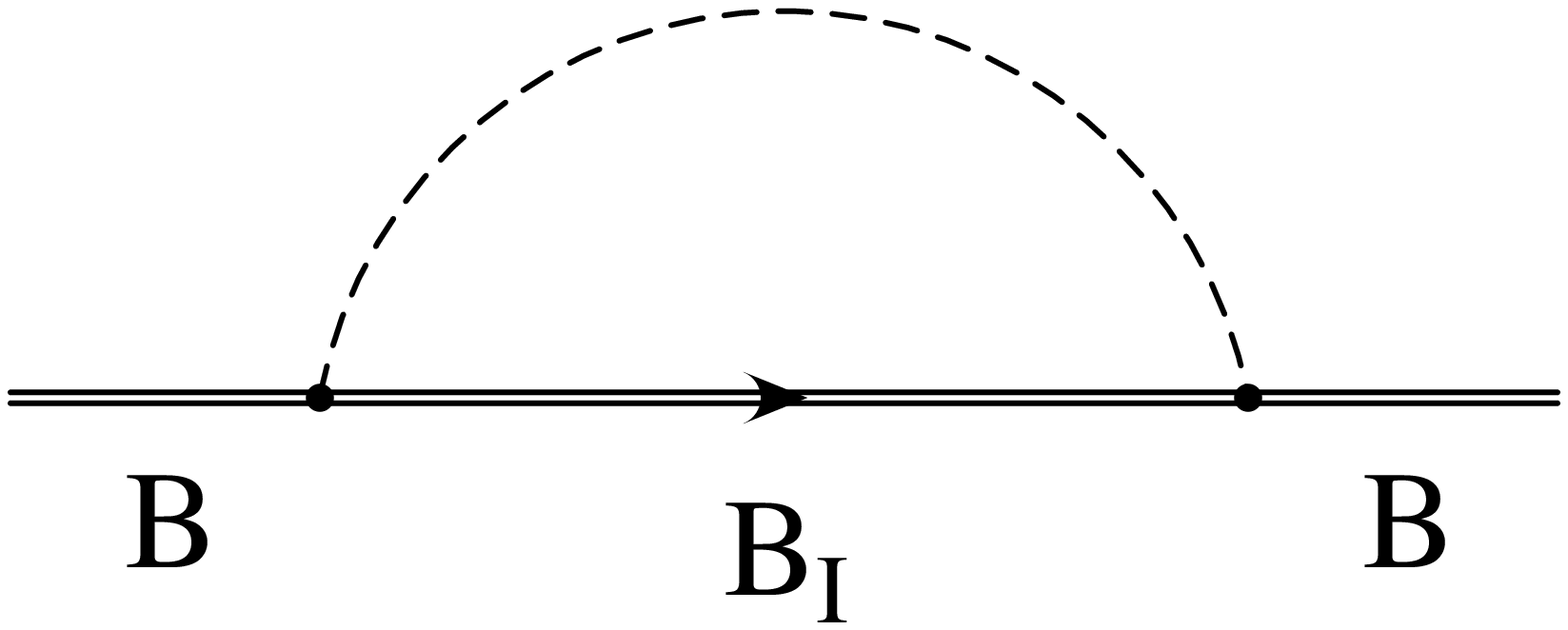}}
\caption{}
\end{figure}
\begin{eqnarray}
-\delta M_i & = & I_{\rm \bf 1} (\pi,K,\eta;\Delta) \, m_{i,{\rm {\bf 1}}} +
I_{\rm \bf 8} (\pi,K,\eta;\Delta) \, m_{i, {\rm {\bf 8}}} + I_{\rm \bf 27}
(\pi,K,\eta;\Delta) \, m_{i, {\rm {\bf 27}}} \, . \label{eq:corr27}
\end{eqnarray}
$I_{\rm \bf j} (\pi,K,\eta;\Delta)$ are flavor lineal combinations of the
integral over the loop, $F(m_\Pi,\Delta)$ \cite{jen96}. Here $m_\Pi$ is the
pion mass and $\Pi= \pi, K, \eta$. Specifically, \begin{eqnarray} &   & I_{\rm
\bf 1} (\pi,K,\eta;\Delta) = \frac18 \left[ 3 F(\pi,\Delta) + 4 F(K,\Delta) +
F(\eta,\Delta) \right], \\  &   & I_{\rm \bf 8} (\pi,K,\eta;\Delta) = \frac{2
\sqrt 3}{5} \left( \frac32 F(\pi,\Delta) - F(K,\Delta) - \frac12
F(\eta,\Delta) \right), \\ &   & I_{\rm \bf 27} (\pi,K,\eta;\Delta) = \frac13
F(\pi,\Delta) - \frac43 F(K,\Delta) + F(\eta,\Delta)
\end{eqnarray}
with
\begin{eqnarray}
\begin{array}{ll}
\displaystyle
m_{i, {\rm {\bf 1}}} = {\bar \lambda}_i^\pi + {\bar \lambda}_i^K +
{\bar \lambda}_i^\eta \, , &  \quad m_{i, {\rm {\bf 8}}} =
\frac{1}{\sqrt 3} \left( {\bar \lambda}_i^\pi - \frac12 {\bar \lambda}_i^K -
{\bar \lambda}_i^\eta \right) \,\\ m_{i, {\rm {\bf 27}}} = \frac{3}{40} \left(
{\bar \lambda}_i^\pi - 3 {\bar \lambda}_i^K + 9 {\bar \lambda}_i^\eta \right)
\,. &  \end{array} \end{eqnarray}

     The $\bar \lambda_i^{\Pi}$ coefficients for the octet baryons read
\begin{eqnarray}
\begin{array}{ll}
\displaystyle
{\bar \lambda}_N^\pi = \frac94 {(F + D)}^2 + 2 {\cal C}^2 \, , \qquad &
\displaystyle
{\bar \lambda}_\Sigma^\pi = 6 F^2 + D^2 + \frac13 {\cal C}^2\, , \\ [3mm]
\displaystyle
{\bar \lambda}_N^K = \frac12 (9 F^2 - 6 F D + 5 D^2 + {\cal C}^2) \, ,
\qquad &
\displaystyle
{\bar \lambda}_\Sigma^K = 3 (F^2 + D^2) + \frac53 {\cal C}^2\, , \\
[3mm]
\displaystyle
{\bar \lambda}_N^\eta = \frac14 {(3 F - D)}^2 \, , \qquad &
\displaystyle
{\bar \lambda}_\Sigma^\eta = D^2 + \frac12 {\cal C}^2 \, , \\ [3mm]
\displaystyle
{\bar \lambda}_\Xi^\pi = \frac94 {(F - D)}^2 + \frac12 {\cal C}^2 \, ,
\qquad &
\displaystyle
{\bar \lambda}_\Lambda^\pi = 3 D^2 + \frac32 {\cal C}^2 \, , \\ [3mm]
\displaystyle
{\bar \lambda}_\Xi^K = \frac12 (9 F^2 + 6 F D + 5 D^2 + 3 {\cal C}^2)
\, , \qquad &
\displaystyle
{\bar \lambda}_\Lambda^K = 9 F^2 + D^2 + {\cal C}^2\, , \\ [3mm]
\displaystyle
{\bar \lambda}_\Xi^\eta =  \frac14 {(3 F + D)}^2 + \frac12 {\cal C}^2
\, , \qquad &
\displaystyle
{\bar \lambda}_\Lambda^\eta = D^2 \, .\label{eq:lreno}
\end{array}
\end{eqnarray}
whereas for the decuplet baryons one has
\begin{eqnarray}
\begin{array}{ll}
\displaystyle
{\bar \lambda}_\Delta^\pi = \frac{25}{36} {\cal H}^2  + \frac12 {\cal C}^2
\, ,  \qquad &
\displaystyle
{\bar \lambda}_{\Xi^*}^\pi = \frac{5}{36} {\cal H}^2 + \frac14 {\cal
C}^2\, , \\ [3mm]
\displaystyle
{\bar \lambda}_\Delta^K = \frac{5}{18} {\cal H}^2 + \frac12 {\cal C}^2 \,
, \qquad &
\displaystyle
{\bar \lambda}_{\Xi^*}^K = \frac{5}{6} {\cal H}^2 + \frac12 {\cal C}^2\, ,
\\ [3mm]
\displaystyle
{\bar \lambda}_\Delta^\eta = \frac{5}{36} {\cal H}^2  \, , \qquad &
\displaystyle
{\bar \lambda}_{\Xi^*}^\eta =  \frac{5}{36} {\cal H}^2 + \frac14 {\cal
C}^2 \, , \\ [3mm]
\displaystyle
{\bar \lambda}_{\Sigma^*}^\pi = \frac{10}{27} {\cal H}^2  + \frac5{12}
{\cal C}^2 \, , \qquad &
\displaystyle
{\bar \lambda}_{\Omega^-}^\pi = \frac{10}{27} {\cal H}^2 \, , \\ [3mm]
\displaystyle
{\bar \lambda}_{\Sigma^*}^K =  \frac{20}{27} {\cal H}^2 + \frac13 {\cal
C}^2 \, , \qquad &
\displaystyle
{\bar \lambda}_{\Omega^-}^K = \frac59 {\cal H}^2  + {\cal C}^2\, , \\
[3mm]
\displaystyle
{\bar \lambda}_{\Sigma^*}^\eta = \frac14 {\cal C}^2 \, , \qquad &
\displaystyle
{\bar \lambda}_{\Omega^-}^\eta =  \frac59 {\cal H}^2 \, + {\cal C}^2. 
\label{eq:lrend}
\end{array}
\end{eqnarray}

    Equation~(\ref{eq:corr27}) can be used to analyze the two flavor-{\bf
27} combinations of baryon masses, namely, the Gell-Mann--Okubo combination for
octet baryons and the equal spacing rule combination for decuplet baryons,
which are respectively
\begin{eqnarray}
\frac34 \Lambda + \frac14 \Sigma - \frac12 (N + \Xi) \, , \qquad -\frac47
\Delta + \frac57 \Sigma^* + \frac27 \Xi^* - \frac37 \Omega \, ,
\end{eqnarray}
where particle labels denote the corresponding masses. These relations are 
a direct consequence of the fact that SU(3) symmetry breaking is purely
octet~\cite{jm255,jm259}.

     The terms proportional to $m_{i, {\rm {\bf 1}}}$ and $m_{i, {\rm {\bf
8}}}$ in Eq.~(\ref{eq:corr27}) have no effect on the above mass relations.
The {\bf 27} piece produces the corrections
\begin{eqnarray}
\left[ -\frac34 (D^2 - 3F^2){\bar I} (0) + \frac18 {\cal C}^2 {\bar I}(\Delta)
\right], \quad \qquad \left[ \frac{5}{18} {\cal H}^2 {\bar I}(0) -\frac14 {\cal
C}^2 {\bar I}(-\Delta) \right] \label{eq:gom}
\end{eqnarray}
to the Gell-Mann--Okubo mass relation and to the equal spacing rule,
respectively, where ${\bar I}(\Delta)$ is an abbreviation for
$I_{\rm \bf 27} (\pi,K,\eta;\Delta)/N_c$ \cite{jen96}. The integral $I_{\rm \bf
27} (\pi,K,\eta;\Delta)$, which is around 4 MeV, is highly suppressed relative
to its singlet and octet counterparts. It explains, however, the small
violation to the Gell-Mann--Okubo mass relation.

\section{Baryons in large $N_c$ chiral perturbation theory}

     The derivation of the $1/N_c$ baryon chiral Lagrangian is given in
Ref.~\cite{jen96}. Recent developments on the large-$N_c$ formalism applied to
baryons can be found in excellent reviews~\cite{manh,Jenkins}. One can also
refer to Jenkins' contribution to these proceedings.

     The $1/N_c$ baryon chiral Lagrangian can be expressed as \cite{jen96}
\begin{eqnarray}
{\cal L_{\rm baryon}} & = & i {\cal D}^0 - M_{\rm hyperfine} + {\rm Tr}
\left( {\cal A}^i \lambda^a \right) A^{ia} + {\rm Tr} \left( {\cal A}^i
{{2I} \over \sqrt{6}} \right) A^{i} + \ldots \label{eq:ncch}
\end{eqnarray}
Each term in Eq.~(\ref{eq:ncch}) involves a baryon
operator which can be given in terms of polynomials in the spin-flavor
generators $J^i$, $T^a$, and $G^{ia}$ \cite{djm95}. For instance,
at the physical value $N_c = 3$, the baryon axial current $A^{ia}$ is
expressed as
\begin{eqnarray}
A^{ia} & = & a_1 G^{ia} + b_2 \frac{1}{N_c} J^i T^a + b_3 \frac{1}{N_c^2}
{\cal D}_3^{ia} + c_3 \frac{1}{N_c^2} {\cal O}_3^{ia} \, ,
\end{eqnarray}
where the operators ${\cal D}_3^{ia}$ and ${\cal O}_3^{ia}$ are defined in
\cite{djm95}.

     We now compute again the leading nonanalytic corrections to the baryon
masses but now within the combined formalism in
$1/N_c$ and chiral corrections. The computation is complicated by the presence
of the hyperfine and quark mass splittings. In the chiral limit $m_i\rightarrow
0$ the baryon propagator is diagonal in spin and can be written as~\cite{jen96}
\begin{eqnarray}
\frac{i {\cal P}_{\sf j}}{ k^0 - \Delta_{\sf j} } \, ,
\label{eq:sproj}
\end{eqnarray}
where ${\cal P}_{\sf j}$ is a spin projection operator for spin $J =
{\sf j}$. For $N_c=3$ one has
~\cite{jen96} 
\begin{eqnarray}
{\cal P}_{1 \over 2} = - {1 \over 3} \left( J^2 - {{15} \over 4}
\right) \, , \qquad {\cal P}_{3 \over 2} = {1 \over 3} \left( J^2 - {3
\over 4} \right) \, .
\end{eqnarray}
On the other hand, $\Delta_{\sf j}$ in Eq.~(\ref{eq:sproj}) stands for
the difference of the hyperfine mass splitting for spin $J = {\sf j}$ and
the external baryon.

     The diagram in Fig.~1, given by the product of a baryon operator
times the pion flavor tensor, can be expressed as
\begin{eqnarray}
{1 \over N_c} \sum_{\sf j} \left( A^{ia} {\cal P}_{\sf j} A^{ib} \right) \
\Pi^{ab} \left(\Delta_{\sf j} \right) \, , \label{eq:massaia}
\end{eqnarray}
where $\Pi^{ab}$ is the symmetric tensor defined as
\begin{eqnarray}
\Pi^{ab} (\Delta) & =  & I_{\rm \bf 1} (\pi,K,\eta;\Delta)\, \delta^{ab} +
I_{\rm \bf 8} (\pi,K,\eta;\Delta)  \, d^{ab8} \nonumber \\
&   & \mbox{} + I_{\rm \bf 27} (\pi,K,\eta;\Delta) \, \Bigl( \delta^{a8}
\delta^{b8} - \frac18 \delta^{ab} - \frac35 d^{ab8} d^{888} \Bigr) \, .
\label{eq:pisym}
\end{eqnarray}
For $N_c = 3$, Eq.~(\ref{eq:massaia}) reduces to
\begin{eqnarray}
{1 \over N_c} \left[ A^{ia} {\cal P}_{1 \over 2} A^{ib} \ \, \Pi^{ab} (
\Delta_{1 \over 2}) + A^{ia} {\cal P}_{3 \over 2} A^{ib} \ \, \Pi^{ab}
(\Delta_{3 \over 2}) \right] \, . \label{eq:massaiash}
\end{eqnarray}

     The evaluation of Eq.~(\ref{eq:massaiash}) involves the computation of
the baryon operators $A^{ia} A^{ib} \Pi^{ab}$ and $A^{ia} J^2 A^{ib}
\Pi^{ab}$. One can follow the approach implemented by Jenkins~\cite{jen96} to
perform the operator reduction of the spin operators involved in the latter
expressions by using spin projection operators. However, we will use a
simplified version of such analysis. After a long but otherwise standard
calculation we find
\begin{eqnarray}
&  & A^{ia} A^{ia} =  \frac{3}{16} N_c(N_c+6) a_1^2 + \frac34 \left( 1 +
 \frac{6}{N_c} \right) a_1 c_3 + \left[- \frac{5}{12} a_1^2 + \frac23 \left(
1 + \frac{3}{N_c} \right) a_1 b_2 \right. \nonumber \\
&   & \mbox{\hglue0.5truecm} + \left.  \left(\frac12 + \frac{3}{N_c} +
\frac{4}{N_c^2} \right) a_1 b_3  + \frac{1}{12} \left(1 + \frac{6}{N_c}
\right) b_2^2 + \left(\frac12 + \frac{3}{N_c} - \frac{9}{N_c^2} \right) a_1
c_3 \right] J^2 \nonumber \\ 
&   & \mbox{\hglue0.5truecm} + \frac{1}{N_c^2} \left[ \frac23 a_1 b_3 + b_2^2
-2 a_1 c_3 +  \frac83  \left(1 + \frac{3}{N_c} \right) b_2 b_3 \right] J^4 +
{\cal O} \left(\frac{1}{N_c^4} \right) \, ,
\end{eqnarray}
\begin{eqnarray}
&   & d^{ab8} A^{ia} A^{ib} =  \left[ \frac38 (N_c+3) a_1^2 + 
\frac{3}{2N_c} \left(1 + \frac{3}{N_c} \right) a_1 c_3 \right] T^8 +
\left[- \frac{7}{12} a_1 ^2 + \frac{3}{N_c^2} a_1 b_3
\right. \nonumber \\
&   & \mbox{\hglue0.5truecm} \left. + \frac16 \left(1 +
\frac{3}{N_c} \right) a_1 b_2 - \frac{9}{2N_c^2} a_1 c_3 \right] \{J^r,
G^{r8}\} + \frac{1}{N_c} \left[ \frac16 a_1 b_2  \right. \nonumber \\
&   & \mbox{\hglue0.5truecm} \left. + \frac12 \left( 1 + \frac{3}{N_c}
\right) \left( a_1 b_3- \frac16 b_2^2 + a_1 c_3 \right) \right] \{J^2, T^8\} +
\frac{1}{N_c^2} \left[-\frac13 a_1 b_3 + \frac12 b_2^2 - a_1 c_3 \right. \nonumber \\
&   & \mbox{\hglue0.5truecm} + \left. \left(
\frac13 + \frac{1}{N_c} \right) b_2 b_3 \right] \{J^2, \{J^r, G^{r8}\}\} +
\frac{2}{3N_c^3} b_2 b_3 \{J^4, T^8\} + {\cal O} \left(\frac{1}{N_c^4} \right)
\, ,
\end{eqnarray}
\begin{eqnarray}
&   & A^{i8} A^{i8} = \frac12 a_1^2 \{G^{i8},G^{i8}\} + \frac{1}{2N_c} a_1
b_2 \{T^8, \{J^i,G^{i8}\}\} + \frac{1}{N_c^2} a_1 b_3
\{\{J^r,G^{r8}\},\{J^i,G^{i8}\}\} \nonumber \\
&   & \mbox{\hglue0.5truecm} + \frac{1}{N_c^2} a_1 c_3 \left[
\{J^2,\{G^{i8},G^{i8}\}\} + [G^{i8},[J^2,G^{i8}]] - \frac12
\{\{J^r,G^{r8}\},\{J^i,G^{i8}\}\} \right] \nonumber \\
&   & \mbox{\hglue0.5truecm} + \frac{1}{4 N_c^2} b_2^2
\{J^2,\{T^8,T^8\}\} + \frac{1}{N_c^3} b_2 b_3 \{J^2,\{T^8,\{J^i,G^{i8}\}\}\} +
{\cal O} \left(\frac{1}{N_c^4} \right) \, .
\end{eqnarray}
Explicit expression for the $T^8$ and $G^{i8}$ operators can be found in
Ref.~\cite{djm95}. Similarly,
\begin{eqnarray}
&   & A^{ia} J^2 A^{ia} = \frac{3}{8} N_c(N_c+6) a_1^2 + \left( \frac32 +
\frac{9}{N_c} \right) a_1 c_3 + \left[ \left( \frac{3}{16}
N_c(N_c+6) -\frac72 \right) a_1^2 \right. \nonumber \\
&   & \mbox{\hglue0.5truecm} + \left. \left( \frac{13}{4} \left(1 + \frac{6}{N_c} 
\right) - \frac{18}{N_c^2} \right) a_1 c_3 \right] J^2  + \left[-\frac{5}{12}
a_1^2 \right. + \frac23 \left(1 + \frac{3}{N_c} \right) a_1 b_2
\nonumber \\
&   & \mbox{\hglue0.5truecm} + \left. \left( \frac12 + \frac{1}{N_c} \right)
\left( 1 + \frac{4}{N_c}\right) a_1 b_3 + \frac{1}{12} \left(1 + \frac{6}{N_c}
\right) b_2^2 + \left(\frac12 + \frac{3}{N_c}  - \frac{27}{N_c^2} \right) a_1
c_3 \right] J^4 \nonumber \\ 
&   & \mbox{\hglue0.5truecm} + \frac{1}{N_c^2} \left[ \frac23 a_1 b_3
+ b_2^2 -2 a_1 c_3  +  \frac83 \left(1 + \frac{3}{N_c} \right) b_2 b_3 \right]
J^6 + {\cal O} \left(\frac{1}{N_c^4} \right) \, ,
\end{eqnarray}
\begin{eqnarray}
&  & d^{ab8} A^{ia} J^2 A^{ib} = \left[ \frac34 (N_c+3) a_1^2 +
\frac{3}{N_c} \left(1 + \frac{3}{N_c} \right) a_1 c_3 \right]T^8 -
\left[2 a_1^2 + \frac{9}{N_c^2} a_1 c_3 \right] \{J^r,
G^{r8}\} \nonumber \\
&   & \mbox{\hglue0.5truecm} + \left[ \frac{3}{16} (N_c+3) a_1^2 + \frac{13}{4} \left(
\frac{1}{N_c} + \frac{3}{N_c^2} \right) a_1 c_3 \right] \{J^2,T^8\} +
\frac{2}{3N_c^3} \{J^6, T^8\} \nonumber \\
&   & \mbox{\hglue0.5truecm} + \left[- \frac{7}{24} a_1^2 +
\frac{1}{12}\left(1 + \frac{3}{N_c} \right) _1 b_2 + \frac{1}{2 N_c^2} a_1
b_3 - \frac{33}{4N_c^2} a_1 c_3 \right] \{J^2,\{J^r, G^{r8}\}\} \nonumber \\
&   & \mbox{\hglue0.5truecm} + \frac{1}{N_c^2} \left[-\frac13 a_1 
b_3 + \frac12 b_2^2 - a_1 c_3 + \frac13 \left( 1
+ \frac{3}{N_c}\right) b_2 b_3 \right] \{J^4, \{J^r, G^{r8}\}\} \nonumber \\
&   & \mbox{\hglue0.5truecm} + \left. \frac{1}{N_c} \left[ \frac16 a_1 b_2
\right. + \frac12 \left( 1 + \frac{3}{N_c} \right) \left( a_1 b_3 - \frac16
b_2^2 + a_1 c_3 \right)  \right] \{J^4, T^8\} +
{\cal O} \left(\frac{1}{N_c^4} \right) \, ,
\end{eqnarray}
\begin{eqnarray}
&  & A^{i8} J^2 A^{i8} = \frac14 a_1^2 \left( \{J^2, \{G^{i8},G^{i8}\}\} +
2 [G^{i8},[J^2,G^{i8}]] \right) + \frac{1}{4N_c} a_1 b_2 \{J^2,\{T^8,
\{J^i,G^{i8}\}\}\} \nonumber \\
&   & \mbox{\hglue0.5truecm} + \frac{1}{2N_c^2} a_1 b_3 \{J^2, \{\{J^r, G^{r8}\}, \{J^i,
G^{i8}\}\}\} + \frac{1}{8 N_c^2} b_2^2 \{J^2,\{J^2,\{T^8,T^8\}\}\}
\nonumber \\
&   & \mbox{\hglue0.5truecm} + \left. \frac{1}{2 N_c^2} a_1 c_3 \right[\{J^2,
\{J^2, \{G^{i8}, G^{i8}\}\}\} + \{J^2,[G^{i8},[J^2,G^{i8}]]\} + 2 [G^{i8},
[J^2, \{J^2, G^{i8}]] \nonumber \\
&   & \mbox{\hglue0.5truecm} \left. - \frac12 \{J^2,\{\{J^r,G^{r8}\},
\{J^i,G^{i8}\}\}\} \right] + \frac{1}{2N_c^3} b_2 b_3
\{J^2,\{J^2,\{T^8,\{J^i,G^{i8}\}\}\}\}  \, . 
\end{eqnarray}

     The full evaluation of the baryon operators leads to
\begin{eqnarray}
&   & \frac34 \Lambda + \frac14 \Sigma - \frac12 (N + \Xi) = \frac{1}{N_c}
\left[ \left(\frac{1}{16} a_1^2 + \frac34 \frac{1}{N_c} a_1 b_2 +
\frac{9}{16} \frac{1}{N_c^2} b_2 ^2 + \frac38 \frac{1}{N_c^2} a_1 b_3
\right. \right. \nonumber \\
&  & \mbox{\hglue1.0truecm} \left. \left. + \frac94 \frac{1}{N_c^3} b_2 b_3
\right) I(0) + \left( \frac18 a_1^2 + \frac98 \frac{1}{N_c^2} a_1 c_3 \right)
I(\Delta) + {\cal O} \left(\frac{1}{N_c^4} \right) \right] \, ,
\end{eqnarray} 
for the Gell-Mann Okubo mass formula and
\begin{eqnarray}
&   & -\frac47 \Delta + \frac57 \Sigma^* + \frac27 \Xi^* - \frac37 \Omega
= \frac{1}{N_c} \left[ \left(\frac{5}{8} a_1^2 + \frac{15}{4} 
\frac{1}{N_c} a_1 b_2 + \frac{45}{8} \frac{1}{N_c^2} b_2 ^2 + \frac{75}{4}
\frac{1}{N_c^2} a_1 b_3 \right. \right. \nonumber \\
&  & \mbox{\hglue1.0truecm} \left. \left. + \frac{225}{4} \frac{1}{N_c^3}
b_2 b_3 \right) I(0) - \left( \frac14 a_1^2 + \frac94 \frac{1}{N_c^2} a_1 c_3
\right) I(-\Delta) + {\cal O} \left(\frac{1}{N_c^4} \right) \right]
\, ,
\end{eqnarray}
for the equal spacing rule. The above equations, at the physical value
$N_c=3$, can be straightforwardly compared with the analogous expressions
obtained in heavy baryon chiral perturbation theory Eqs.~(\ref{eq:gom})
by using the identifications
\begin{eqnarray}
\begin{array}{l}
\displaystyle
D = \frac12 a_1 + \frac16 b_3 \, , \\ [3mm]
\displaystyle
F = \frac13 a_1 + \frac16 b_2 + \frac19 b_3 \, , \\ [3mm]
\end{array}
\quad \quad
\begin{array}{l}
\displaystyle
{\cal C} = - a_1 - \frac12 c_3\, , \\ [3mm]
\displaystyle
{\cal H} = - \frac32 a_1 - \frac32 b_2 - \frac52 b_3 \, , \\ [3mm]
\end{array} \label{eq:cid}
\end{eqnarray}

    Both expressions agree.

\section{Conclusions}
We have exemplified how to compute corrections to baryon masses in a
calculational scheme that simultaneously exhibits both the $m_q$ and $1/N_c$
expansions. Flavor-$\bf 27$ baryon mass splittings at leading order in chiral
perturbation theory have been computed in detail to illustrate the procedure.
Furthermore, our results for three colors have been compared with
the ones obtained in heavy baryon chiral perturbation theory with no $1/N_c$
corrections and an agreement has been found term by term in the series. 

     Some other applications of the combined approach to baryon properties will
be presented elsewhere.

\begin{theacknowledgments}
The author is grateful to Consejo Nacional de Ciencia y Tecnolog{\'\i}a
(Mexico) for partial support.
\end{theacknowledgments}

\end{document}